\documentstyle[sprocl,epsfig]{article}

\def\be{\begin{equation}}
\def\ee{\end{equation}}
\def\bea{\begin{eqnarray}}
\def\eea{\end{eqnarray}}

\newcommand{\case}[2]{\mbox{\footnotesize $\displaystyle \frac{#1}{#2}$}}
\newcommand{\lsim}{\mathrel{\rlap{\lower3pt\hbox{\hskip0pt$\sim$}}
\raise2pt\hbox{$<$}}}
\newcommand{\gsim}{\mathrel{\rlap{\lower3pt\hbox{\hskip0pt$\sim$}}
\raise2pt\hbox{$>$}}}

\begin{document}

\title{HEAVY MESON OBSERVABLES}

\author{M. A. IVANOV,\footnotemark[1] 
Yu. L. KALINOVSKY$\,$\footnotemark[2]
and C. D. ROBERTS$\,$\footnotemark[3]}

\address{
\footnotemark[1]Bogoliubov Laboratory of Theoretical Physics,
 \\Joint Institute for Nuclear Research, 141980 Dubna,
 Russia\vspace*{0.2\baselineskip}\\  
 \footnotemark[2]Laboratory of Computing Techniques and Automation, \\Joint
 Institute for Nuclear Research, 141980 Dubna,
 Russia\vspace*{0.2\baselineskip}\\  
\footnotemark[3]Physics Division, Bldg. 203, Argonne National Laboratory\\
Argonne IL 60439-4843, USA}


\maketitle\abstracts{An accurate and unified description of the non-hadronic
electroweak interactions of light- and heavy-mesons is possible when the
dressing of quark propagators and the finite size of hadronic bound states is
explicitly accounted for.}

The study$\,$\cite{ivanov} of non-hadronic electroweak interactions is an
excellent way to learn more about the long-range behaviour of the QCD
interaction because the probes are very well understood.  This is
important$\,$\cite{robertsect} because only with a quantitative understanding
of this part of the interaction can the properties of the deconfinement and
chiral symmetry restoration transitions be determined.

Light- {\it and} heavy-mesons are bound states of a dressed-quark and
-anti\-quark.  That dressing is described$\,$\cite{cdragw} by the quark
Dyson-Schwinger equation (DSE).  Heavy quarks are distinguished by their
almost momentum-indepen\-dent mass function, which suggests the approximation:
\begin{equation}
\label{dsehq}
S^{-1}_Q(p) = i\gamma \cdot p + \hat M_Q\,,\; Q=c,b\,,
\end{equation}
where $\hat M_Q$ is a constituent-heavy-quark mass parameter.  This behaviour
is in marked contrast to the significant momentum dependence$\,$\cite{ivanov}
of light $u$-, $d$-, $s$-quark propagators.  

\begin{table}[t]
\hspace*{-7.35pt}
\begin{tabular}{lll|lll}
        & Obs.  & Calc. & & Obs.  & Calc. \\\hline
$f_\pi$   & 0.131 & 0.146 & $m_\pi$   & 0.138 & 0.130 \\
$f_K  $   & 0.160 & 0.178 & $m_K$     & 0.496 & 0.449 \\
$\langle \bar u u\rangle^{1/3}$ & 0.241 & 0.220 &
        $\langle \bar s s\rangle^{1/3}$ & 0.227 & 0.199\\
$\langle \bar q q\rangle_\pi^{1/3}$ & 0.245 & 0.255& 
        $\langle \bar q q\rangle_K^{1/3}$ & 0.287 & 0.296\\
$f_\rho$   & 0.216      & 0.163 & 
        $f_{K^\ast}$   & 0.244      & 0.253 \\
$\Gamma_{\rho\pi\pi}$ & 0.151     & 0.118 & 
        $\Gamma_{K^\ast (K\pi)}$  & 0.051     & 0.052 \\
$f_D$   & 0.200 $\pm$ 0.030     & 0.213 & 
        $f_{D_s}$ & 0.251 $\pm$ 0.030     & 0.234 \\
$f_B$   & 0.170 $\pm$ 0.035      & 0.182 & 
$g_{B K^\ast \gamma} \hat M_b$ & 2.03 $\pm$ 0.62 & 2.86 \\[2pt]\hline
$f_+^{B\to D}(0)$ & 0.73 & 0.58  &
        $f_\pi r_\pi$ & 0.44 $\pm$ 0.004 & 0.44   \\
$F_{\pi\,(3.3\,{\rm GeV}^2)}$ & 0.097 $\pm$  0.019  & 0.077 &
        B$(B\to D^\ast)$ & 0.045 $\pm$ 0.003 & 0.052\\
$\rho^2$ &  1.53 $\pm$ 0.36 & 1.84 &
        $\alpha^{B\to D^\ast}$ & 1.25 $\pm$ 0.26 & 0.94 \\
$\xi(1.1)$  & 0.86 $\pm$ 0.03& 0.84 &
        $A_{\rm FB}^{B\to D^\ast}$ & 0.19 $\pm$ 0.031 & 0.24 \\
$\xi(1.2)$  & 0.75 $\pm$ 0.05& 0.72 &
        B$(B\to \pi)$ & $1.8 \pm 0.6_{\times 10^{-4}}$  & 2.2 \\
$\xi(1.3)$  & 0.66 $\pm$ 0.06& 0.63 &
        $f^{B\to \pi}_{+(14.9\,{\rm GeV}^2)}$ & 0.82 $\pm$ 0.17 & 0.82 \\
$\xi(1.4) $ & 0.59 $\pm$ 0.07& 0.56 &
        $f^{B\to \pi}_{+(17.9\,{\rm GeV}^2)}$ & 1.19 $\pm$ 0.28 & 1.00 \\
$\xi(1.5) $ & 0.53 $\pm$ 0.08& 0.50 &
        $f^{B\to \pi}_{+(20.9\,{\rm GeV}^2)}$ & 1.89 $\pm$ 0.53 & 1.28 \\
B$(B\to D)$ & 0.020 $\pm$ 0.007 & 0.013 &
        B$(B\to \rho)$ & $2.5 \pm 0.9_{\times 10^{-4}}$ & 4.8 \\
B$(D\to K^\ast)$ & 0.047 $\pm$ 0.004  & 0.049 &
        $f_+^{D\to K}(0)$ & 0.73 &  0.61 \\
$\frac{V(0)}{A_1(0)}^{D \to K^\ast}$ & 1.89 $\pm$ 0.25 & 1.74 &
        $f_+^{D\to \pi}(0)$ & 0.73 &  0.67 \\
$\frac{\Gamma_L}{\Gamma_T}^{D \to K^\ast}$ & 1.23 $\pm$ 0.13 & 1.17 &
        $g_{B^\ast B\pi}$ & 23.0 $\pm$ 5.0 & 23.2 \\
$\frac{A_2(0)}{A_1(0)}^{D \to K^\ast}$ & 0.73 $\pm$ 0.15 & 0.87 &
        $g_{D^\ast D\pi}$ & 10.0 $\pm$ 1.3 & 11.0 \\[2pt]\hline
\end{tabular}
\caption{The 16 dimension-GeV (upper section) and 26 dimensionless (lower
section) quantities used in $\chi^2$-fitting the parameters.  The ``Obs.''
column contains other calculations or experimental
results$\,$\protect\cite{mr97,flynn,pdg98}$^-$\protect\cite{gHsHpi}.\hspace*{\fill}
\label{tablea} }
\end{table}

Bound states are described by a Bethe-Salpeter amplitude, which characterises
the restrictions on the relative momentum of the constituents.  This means
that a further approximation to Eq.~(\ref{dsehq})
\begin{equation}
\label{hqf}
S_Q(k+P) = \case{1}{2}\,\frac{1 - i \gamma\cdot v}{k\cdot v - E_H}
+ {\rm O}\left(|k|/\hat M_{Q},E_H/\hat M_{Q}\right)
\end{equation}
where $P_\mu=: m_H v_\mu=: (\hat M_Q + E_H)v_\mu$, $m_H$ is the hadron's mass
and $k$ is the momentum of the lighter constituent, can only be reliable if
{\it both} the momentum-space width of the Bethe-Salpeter amplitude,
$\omega_H$, and the binding energy, $E_H$, are significantly less than $\hat
M_Q$.

The DSE framework reproduces all of the acknowledged consequences of
heavy-quark symmetry and predicts that the mass of a heavy-meson rises
linearly with the current-mass of its heaviest constituent.\cite{ivanov} This
last result follows from a simple formula, exact in QCD,$\,$\cite{mr97,mrt}
which also yields the Gell-Mann--Oakes--Renner relation in the limit of small
current-quark masses and hence unifies the heavy- and light-quark extremes.

Earlier work$\,$\cite{cdragw} provides the foundation for a phenomenological
application of the DSEs to an extensive but not exhaustive body of
observables:$\,$\cite{ivanov} heavy-meson leptonic decays, semileptonic
heavy-to-heavy and heavy-to-light transitions - $B \to D^\ast$, $D$, $\rho$,
$\pi$; $D \to K^\ast$, $K$, $\pi$, radiative and strong decays -
$B_{(s)}^\ast \to B_{(s)}\gamma$; $D_{(s)}^*\to D_{(s)}\gamma$, $D \pi$, and
the rare $B\to K^\ast \gamma$ flavour-changing neutral-current process.  A
characterisation of the dressed-quark propagators and bound state amplitudes
employing ten parameters, plus the four quark masses, is used in a
$\chi^2$-fit to $N_{\rm obs}=42$ heavy- and light-meson observables.

This yields current-quark masses 
$
m_{u,d} = 5.4\,{\rm MeV},\; m_s = 119\,{\rm MeV},
$
consti\-tuent-quark masses
$
\hat M_c = 1.32\,{\rm GeV},\; \hat M_b = 4.65\,{\rm GeV},
$
and $\omega_H=1.81\,$GeV for the width of the Gaussian {\it Ansatz} for the
heavy-meson Bethe-Salpeter amplitude, with $\chi^2/{\rm d.o.f} = 1.75$ and
$\chi^2/N_{\rm obs} = 1.17$.  The quality of the fit is illustrated in
Table~1.

The results indicate that the heavy-meson binding energy is large, which
means that in current experiments there is a signficant deviation from the
predictions of heavy-quark symmetry, even in many ratios of observables.  The
deviations are $\lsim 30$\% in $b\to c$ transitions and can be as large as a
factor of 2 in $c\to d$ transitions.  Further, the demonstrated
phenomenological efficacy of Eq.~(\ref{dsehq}) and the more sophisticated
description required for the light-quark propagator provides
semi-quantitative constraints$\,$\cite{mr97} on the behaviour of the QCD
interaction at large distances.

\hspace*{-\parindent}{\bf Acknowledgments}.~We thank the staff of the
ECT$^\ast$ for their hospitality and support.  This work was supported in
part by the Russian Fund for Fundamental Research, under contract nos.\
97-01-01040 and 99-02-17731-a, and the US Department of Energy, Nuclear
Physics Division, under contract no.\ W-31-109-ENG-38.

\begin{flushleft}

\end{flushleft}

\end{document}